\documentclass{aastex63}
\pdfoutput=1

\usepackage{xcolor}

\received{April 17, 2020}
\accepted{July 1, 2020}
\shorttitle{Lopsided Satellite Distributions}
\shortauthors{Brainerd \& Samuels}

\begin{document}

\title{Lopsided Satellite Distributions around Isolated 
Host Galaxies}

\correspondingauthor{Tereasa G. Brainerd}
\email{brainerd@bu.edu}

\author[0000-0001-7917-7623]{Tereasa G. Brainerd}
\affil{Boston University \\
Department of Astronomy \& Institute for Astrophysical Research\\
725 Commonwealth Avenue \\
Boston, MA 02215 USA}

\author{Adam Samuels}
\affil{Boston University \\
Department of Astronomy \& Institute for Astrophysical Research\\
725 Commonwealth Avenue \\
Boston, MA 02215 USA}



\begin{abstract}
We investigate the spatial distribution of the satellites of bright, isolated
host galaxies.  
\textcolor{black}{In agreement with previous studies, we find that,} on
average, the satellites of red hosts are
found preferentially close to their hosts' major axes, while the satellites
of blue hosts are distributed 
 isotropically.  
We compute the pairwise clustering of the satellites and 
find a strong tendency 
for pairs of satellites
to be located on the same side of their host, resulting in lopsided
\textcolor{black}{spatial} distributions.  The signal is most pronounced for
the satellites of blue hosts, where 
\textcolor{black}{the number of} pairs
on the same side of their host 
\textcolor{black}{exceeds the number of pairs} on opposite sides of their
host \textcolor{black}{by a factor of $1.8 \pm 0.1$}.  
\textcolor{black}{
For the satellites of red hosts, the number of pairs on the same side
of their host exceeds the number of pairs on opposite sides
of their host by a factor of $1.08 \pm 0.03$.}
\textcolor{black}{
Satellites that are far from their hosts
($r_p \gtrsim 300$~kpc) show a strong preference for
being located on the same side of their hosts; satellites that are near to their hosts
($r_p \lesssim 100$~kpc) show a weak preference for being located on
opposite sides of their hosts.}
While lopsided distributions 
have been found previously 
for the satellites of \textcolor{black}{bright pairs of} galaxies, 
ours is the first study to find lopsided distributions 
for the satellites of isolated bright galaxies.

\end{abstract}



\section{Introduction} \label{sec:intro}


Studies of the spatial distributions of the satellites of isolated
host galaxies have \textcolor{black}{a considerable} history 
\textcolor{black}{
(e.g., Sales \&
Lambas 2004, 2009;
Brainerd 2005; Azzaro et al.\ 2007; Bailin et al.\ 2008; 
\'Ag\'ustsson \&
Brainerd 2010, hereafter AB10; 
\'Ag\'ustsson \& Brainerd 2011).}
The primary motivation
of these studies is a desire to use the satellites as probes of their
hosts' dark matter halos.  In particular, the spatial distributions
of the satellites might provide insight into 
the shapes and assembly
histories of the hosts' dark matter halos.  Constraints on
the orientations
of the host galaxies within their halos \textcolor{black}{might
also} be obtained 
\textcolor{black}{from the locations of satellite galaxies
(AB10).}
Generally, only one or two 
satellites are found \textcolor{black}{for each host} when host-satellite samples
are selected using redshift space criteria (i.e., proximity to each other in both
line-of-sight velocity and projected distance on the sky).
This is \textcolor{black}{often} attributed to the bright
limiting magnitudes of the redshift surveys used to obtain 
the samples; e.g., the Sloan Digital
Sky Survey (SDSS; Fukugita et al.\ 1996; Hogg et al.\ 2001; Smith et al.\
2002; Strauss et al. 2002; York et al.\ 2000) and the Two Degree Field
Galaxy Redshift Survey (2dFGRS; Colless et al.\ 2001, 2003).
When \textcolor{black}{each host has}
only one or two satellites, this necessarily
limits the degree to which statistics can be performed.

\textcolor{black}{Previous studies have shown that
the satellites of pairs of host galaxies
(i.e., large, similarly-bright galaxies that
are found relatively close to each other) are distributed 
asymmetrically with respect to the 
individual hosts. For example,
Conn et al.\ (2013) found that $\sim 80$\% of
M31's satellites are located between M31 and the
Milky Way.  Additionally,
Libeskind et al.\ (2016) studied pairs of bright SDSS galaxies and
their satellites, finding that the satellites were, on average, 
located between the
hosts, rather than in symmetric distributions centered on
each host.  Such ``lopsided'' satellite distributions for 
pairs of bright galaxies have also been found in $\Lambda$ Cold Dark 
Matter simulations (Pawlowski et al.\ 2017; Gong et al.\ 2019), and
they have been attributed to satellites that are on their first approaches
to their hosts (Gong et al.\ 2019). In the case of isolated host galaxies,
systematic lopsidedness of the satellite locations has not been
previously studied.
}

\textcolor{black}{Here} we present results from a new analysis of the spatial
locations of the satellites of isolated host galaxies, selected 
using typical redshift space criteria.  
\textcolor{black}{Each of our host-satellite systems}
\textcolor{black}{is required to have} at least \textcolor{black}{two} satellites,
\textcolor{black}{and this is a}
key difference between our sample and previous samples
that were selected using similar criteria.
\textcolor{black}{The requirement of at least two satellites per 
system allows us
to quantify the locations of the satellites with respect to 
each other via a pairwise clustering statistic.}
The 
properties of the host-satellite sample, and the
criteria used to select it,
are discussed in \S2.  The locations of the satellite galaxies,
measured relative to the major axes of their hosts, and the clustering of the satellites
with each other, are presented in \S3.  The main results are summarized 
and discussed in \S4.
Throughout, we adopt values for the fundamental
cosmological parameters of $H_0 = 70$~km~s$^{-1}$~Mpc$^{-1}$, 
$\Omega_\Lambda = 0.7$,
and $\Omega_{m0} = 0.3$. \textcolor{black}{Below we quote
1$\sigma$ error bars, obtained from 1,000 bootstrap resamplings
of the dataset. Error bars are omitted from figures when they
are comparable to or smaller than the points in
the figures.}

\section{Host-Satellite Sample} \label{sec:sample}

Isolated host galaxies and their satellites were selected from the
NASA-Sloan Atlas (NSA) catalog v1\_0\_1.  The NSA catalog is
publicly available via the
{\tt nsatlas} Table in the SDSS 
database.
The NSA contains virtually all galaxies with spectroscopic redshifts
$z < 0.15$ within the footprint of the 11$^{\rm th}$ data release of the SDSS,
and it incorporates a background subtraction method that
yields significantly improved photometry for galaxies that subtend large angles
on the sky (see Blanton et al.\ 2011).  Throughout, we restrict our
analysis to galaxies with extinction corrected magnitudes
$r < 17.77$, consistent with the SDSS spectroscopic
completeness limit.  Candidate host galaxies were 
required to have \textcolor{black}{line-of-sight} velocities $v > 500$~km~s$^{-1}$
and \textcolor{black}{to} be at least 1~magnitude brighter than all other galaxies
within a projected distance $r_p < 700$~kpc and 
\textcolor{black}{line-of-sight} velocity 
difference $|dv| < 1,000$~km~s$^{-1}$.  
Further, candidate host galaxies
were required to have apparent magnitudes in the range
$9 < r < 15$, insuring that the vast majority of the hosts 
(\textcolor{black}{91.5}\%) have
luminosities brighter than $L_\ast$ (i.e., $M_{r,*} = -20.83$; Blanton
et al.\ 2001).

Candidate satellite galaxies were required to be within a 
projected distance $r_p < 500$~kpc of their hosts, and
the host-satellite \textcolor{black}{line-of-sight} velocity difference was required to be
$|dv| < 500$~km~s$^{-1}$.  Because we are interested in 
studying 
\textcolor{black}{the pairwise clustering of the satellites,}
we \textcolor{black}{further} restricted the sample to 
\textcolor{black}{only} those
systems for which \textcolor{black}{two} or more satellites were found. 
We eliminated \textcolor{black}{344 candidate} systems 
due to various issues that were revealed
upon visual inspection.  \textcolor{black}{These issues included stars
that were incorrectly listed as galaxies in the NSA, galaxies that
were listed twice in the NSA, and systems in which a given candidate
satellite was matched with more than one candidate host.}
Our final, \textcolor{black}{primary}  sample
consists of \textcolor{black}{3,575} host-satellite systems 
with a total of \textcolor{black}{13,090}
satellites.  \textcolor{black}{Below we refer to this
as ``sample~1''.} \textcolor{black}{Various}
properties of sample~1 \textcolor{black}{are summarized in
Fig.\ 1.} 

\begin{figure}[ht!]
\epsscale{0.85}
\plotone{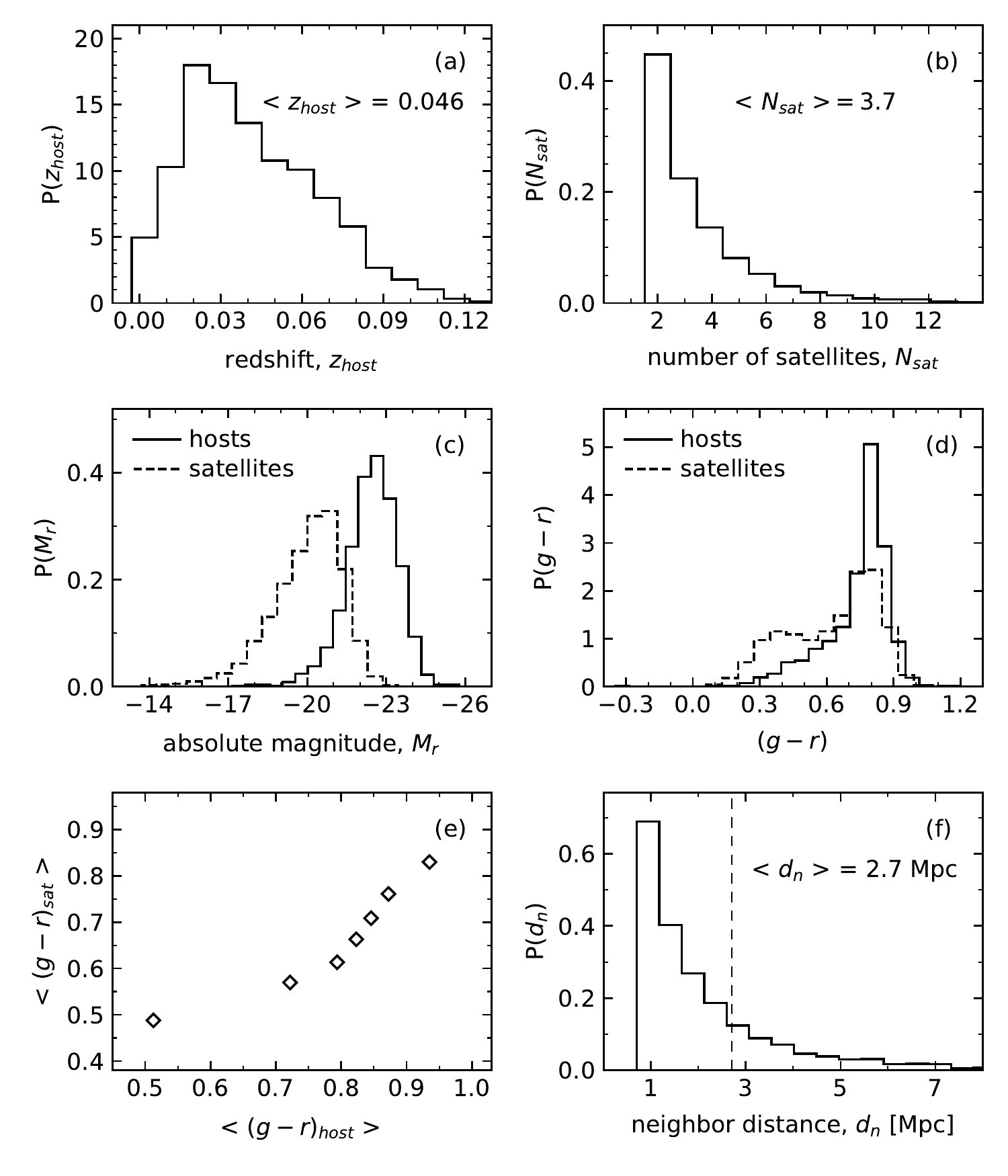}
\caption{
Properties of \textcolor{black}{sample~1}: a) Host redshift distribution. 
b) Number of satellites per host, truncated for clarity; 35
systems have more than 14 satellites. c) Distributions of $r$-band absolute 
magnitudes. 
d) Restframe $(g-r)$ color distributions.
e) Dependence of \textcolor{black}{mean} satellite color
on host color. f) \textcolor{black}{Distribution of projected distances to the
nearest, similarly-bright NSA neighbor galaxy for each host (see text), truncated
for clarity; 165 hosts have $d_n > 8$~Mpc}.
}
\label{fig:properties}
\end{figure}

The total number of satellites per system ranges from \textcolor{black}{2}
to 26, the mean number of satellites per system is \textcolor{black}{3.7}, 
and the median
number of satellites per system is \textcolor{black}{3}.  The median host redshift
is $z_{\rm med} = \textcolor{black}{0.042}$, the median host absolute magnitude
(extinction- and k-corrected) is $M_r = \textcolor{black}{-22.8}$, the median satellite
absolute magnitude is $M_r = \textcolor{black}{-19.9}$, and the mean host-to-satellite
luminosity ratio is \textcolor{black}{19.9} 
(i.e., on average the satellites are 
\textcolor{black}{2.7} 
magnitudes fainter than their hosts).
Below, we subdivide the sample by 
$(g-r)$ color (extinction- and k-corrected), where we take galaxies
with $(g-r) < 0.7$ to be ``blue'' and galaxies with $(g-r) \ge 0.7$
to be ``red'' (e.g., Blanton et al.\ 2003).  The hosts are
predominately red (\textcolor{black}{2,719} hosts)
\textcolor{black}{and approximately}
\textcolor{black}{half} of
the satellites are red (\textcolor{black}{6,873} satellites).
Fig.\ 1e) shows the expected strong color-color correlation of the host
galaxies with their satellites; i.e., \textcolor{black}{on average,}
the satellites of blue hosts are 
\textcolor{black}{typically} blue and the satellites of red hosts are 
\textcolor{black}{typically}
red. \textcolor{black}{We also note that the mean satellite color 
is
essentially independent of host-satellite separation, with
$\left< \left( g - r \right) \right> = 0.501 \pm 0.004$ for
the satellites of blue hosts and
$\left< \left( g - r \right) \right> = 0.698 \pm 0.002$ for
the satellites of red hosts.}

\textcolor{black}
{
To explore the degree to which the hosts in sample~1 can be 
considered to be truly isolated, we computed the distance
between
each host and the nearest NSA galaxy for which the
line-of-sight velocity difference 
was $|dv| < 1,000$~km~s$^{-1}$,
and for which the host was {\it not} at least 1~magnitude brighter.  
These are objects that would cause any
candidate host to fail the above isolation criteria if they were within
700~kpc of the host. 
The distribution of the distances,
$d_n$, between
these objects and the host galaxies is shown in Fig.\ 1f).  From this,
it is clear that the nearest, similarly-bright NSA neighbor galaxies are typically
more than 1~Mpc from the hosts, with a mean distance of
2.7~Mpc.  In addition, only 179 hosts in sample~1 would have passed
the Libeskind et al.\ (2016) selection criteria for bright galaxy pairs,
so we conclude that sample~1 is not significantly contaminated by 
galaxy pairs.  Below, we will 
identify systems that are even more isolated
than sample~1 by imposing additional restrictions, 
and we will explore the effects of these additional 
restrictions on our results for satellite galaxy locations.
}

\section{Locations of Satellite Galaxies} \label{sec:results}

To compare results for the locations of the satellites in 
\textcolor{black}{sample~1} to
results from previous, \textcolor{black}{similar} samples,
we compute \textcolor{black}{their} locations,
$\phi$, measured with respect to \textcolor{black}{their} hosts' major axes.  
\textcolor{black}{Here},
$\phi$ is a polar angle defined such that satellites with 
$\phi = 0^\circ$ are located along the 
direction of the host's
major axis and satellites with $\phi = 90^\circ$ are located 
along the direction of the host's minor axis.
\textcolor{black}{Averaged} over the entire
sample, the locations of the satellites are anisotropic, with a preference
for being found near the major axes of their hosts: $\left< \phi \right> =
\textcolor{black}{42.88^\circ \pm 0.23^\circ}$.  
On average,
the satellites of red hosts are distributed anisotropically with 
$\left< \phi \right> = \textcolor{black}{42.34^\circ \pm 0.25^\circ}$, 
and the satellites of
blue hosts are consistent with an isotropic distribution: $\left< \phi \right> =
\textcolor{black}{44.61^\circ \pm 0.52^\circ}$.  
On average, red satellites are distributed
anisotropically with $\left< \phi \right> = 
\textcolor{black}{42.01^\circ \pm 0.32^\circ}$, 
\textcolor{black}{and}
blue satellites are distributed 
\textcolor{black}{approximately} isotropically:
$\left< \phi \right> = 
\textcolor{black}{44.07^\circ \pm 0.33^\circ}$.  
\textcolor{black}{These results are in excellent
agreement with previous studies of the locations of the satellites
of isolated host galaxies (see, e.g., AB10).
}


\begin{figure}[t!]
\epsscale{1.25}
\plotone{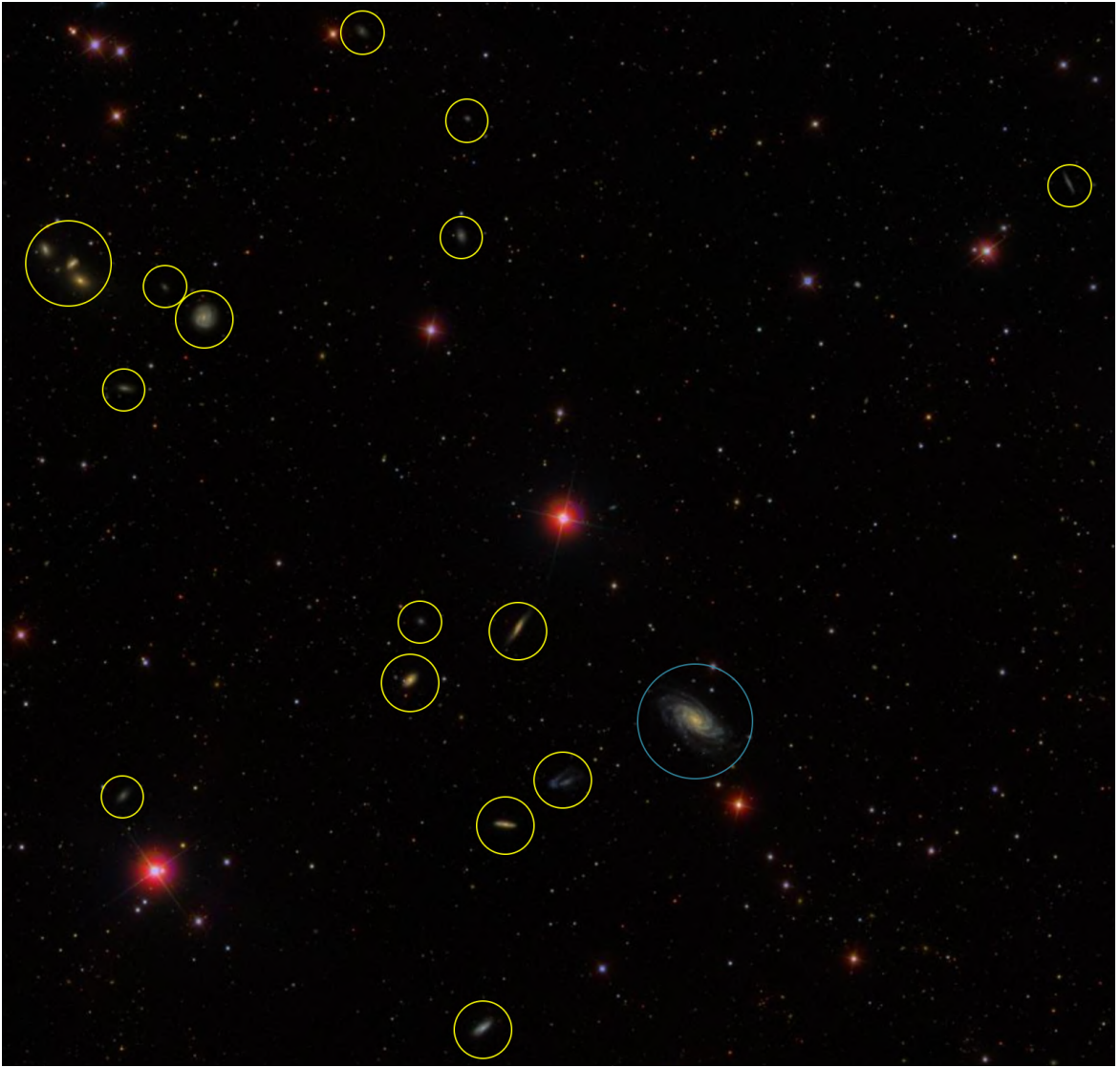}
\caption{
\textcolor{black}{NGC2998 (blue circle) and its 17 satellites (yellow circles).}
}
\label{fig:ngc2998}
\end{figure}


During the process of visually inspecting all host-satellite systems, 
it became clear that many systems had satellite distributions that 
were \textcolor{black}{obviously lopsided}
with respect to the host.  Fig.\ \textcolor{black}{2} shows an example
of one such system.  The host, NGC 2998
(UGC 5250), is a spiral \textcolor{black}{galaxy at} redshift $z = 0.016$. NGC~2998 has
17 satellites, 
\textcolor{black}{16 of which are located to the east of NGC2998 and 12
which are located in the quadrant of the sky to the northeast of NGC2998.}
To determine whether such lopsidedness in the satellite
locations was present throughout our sample,
we constructed a probability distribution for the polar angle differences
between pairs of satellites, $P(\Delta \phi$).  We define $\Delta \phi$
such that pairs of satellites 
with $\Delta \phi \sim 0^\circ$ are
located on the same side of their host, and pairs of satellites with 
$\Delta \phi \sim 180^\circ$ are located on opposite sides of their host.
\textcolor{black}{The top panel of} Fig.\ \textcolor{black}{3} shows an illustration 
of our definition of $\Delta \phi$.
\textcolor{black}{The bottom panel of Fig.\ 3 shows the probability distribution,
$P(\Delta \phi)$, that would result if the satellite locations 
were drawn from a uniform
elliptical distribution.  Uniform elliptical distributions yield
functions for $P(\Delta \phi)$ that are symmetric about $\Delta \phi =
90^\circ$.  The more elliptical the distribution (i.e., the smaller the 
axis ratio, $b/a$), the greater is the value of $P(\Delta \phi)$ for both
$\Delta \phi \sim 0^\circ$ and $\Delta \phi \sim 180^\circ$.  
In the case of a uniform circular distribution, all values
of $\Delta \phi$ are equally probable.
}

\begin{figure}[t!]
\epsscale{0.85}
\plotone{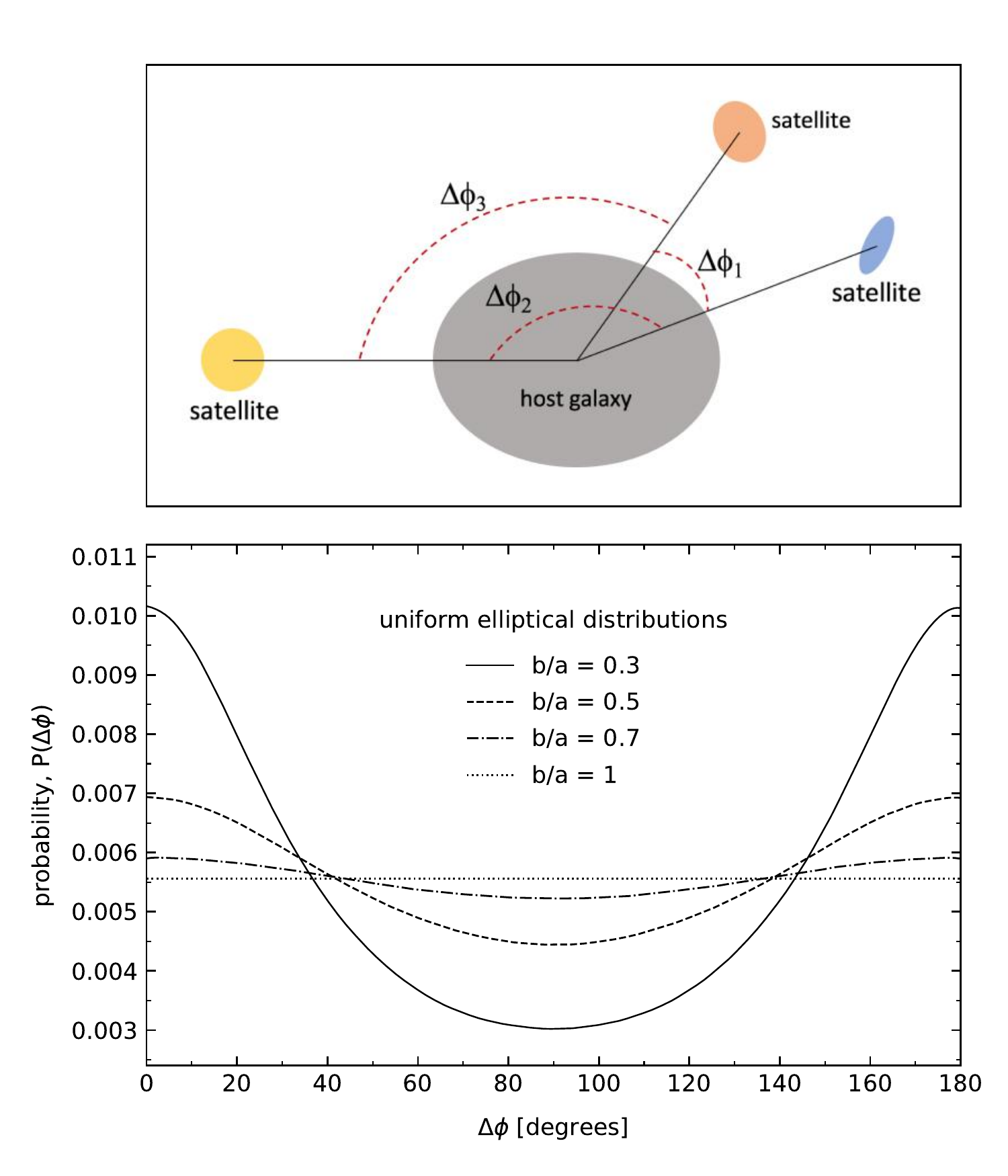}
\caption{
\textcolor{black}{{\it Top:} Illustration of the polar angle difference between
pairs of satellites, $\Delta \phi$. 
{\it Bottom:} Probability distributions, $P(\Delta \phi)$,
expected for uniform elliptical distributions with a range of
axis ratios, $b/a$.
}
}
\label{fig:cartoon}
\end{figure}

Fig.\ \textcolor{black}{4} shows $P(\Delta \phi)$ for 
\textcolor{black}{several subdivisions of sample~1, from which}
it is clear that there are \textcolor{black}{many
more pairs of satellites on the same side of their host
than there are on opposite sides. The effect is most pronounced 
for the satellites of blue hosts and for systems that
have only 2 or 3 satellites. We quantify this by computing the
ratio, $f$, of the number of 
pairs of satellites with $0^\circ \le \Delta \phi \le 20^\circ$ 
and the number of pairs of satellites with
$160^\circ \le \Delta \phi \le 180^\circ$. For the satellites of
blue hosts, $f_{\rm blue} = 1.8 \pm 0.1$, while for the satellites of
red hosts $f_{\rm red} = 1.08 \pm 0.03$.  For the systems with
2 or 3 satellites, $f_{2-3} = 1.34  \pm 0.08$, while for the systems
with 7 to 26 satellites $f_{7-26} = 1.10 \pm 0.03$. 
}

We \textcolor{black}{use $\chi^2$ tests to}
compare $P(\Delta \phi)$ for the satellites to 
$P(\Delta \phi)$ \textcolor{black}{for
a random distribution.}  
\textcolor{black}{To do this, we generate 1,000 independent
random realizations of the satellite polar angles and 
we use these to compute the corresponding
$P(\Delta \phi)$.  
Results for $P(\Delta \phi)$ from the randomized
satellite polar angles are shown
in Fig.\ 4 (circles).
}
Values of the reduced $\chi^2$ (i.e., the $\chi^2$ per degree
of freedom, $\chi^2/\nu$) are shown in each panel of 
Fig.\ \textcolor{black}{4}.  The $\chi^2$
\textcolor{black}{tests reject} the null hypothesis (i.e., that 
$P(\Delta \phi)$ for the satellites is drawn from 
\textcolor{black}{the randomized distribution) with a confidence
level $\ge 99.9999\%$ in all cases.}
To further compare
the clustering of the satellites to that of \textcolor{black}{the 
randomized distribution},
we perform Kolmogorov-Smirnov (KS) tests for
\textcolor{black}{ $P(< \Delta \phi)$, the cumulative
probability distributions for $\Delta \phi$.}
The KS tests reject the null hypothesis
\textcolor{black}{with a confidence level $> 99.9999$\% in all cases.}
We therefore conclude that the satellites are \textcolor{black}{considerably}
more clustered \textcolor{black}{in polar angle}
than would be expected in a \textcolor{black}{randomized} distribution, with pairs
of satellites having a preference for being located on the same side of their
host.

\begin{figure}[t!]
\epsscale{1.15}
\plotone{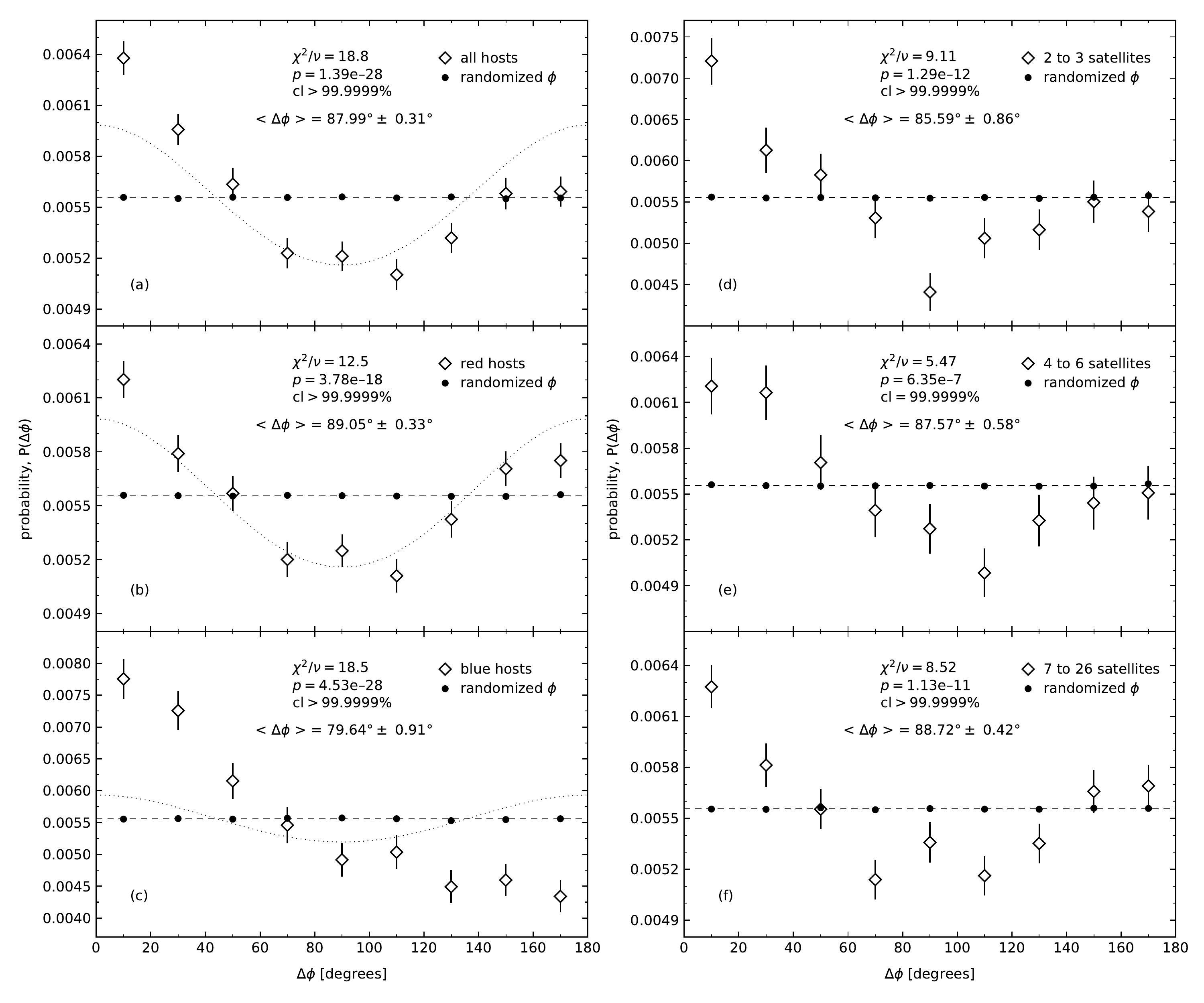}
\caption{
Probability distributions, $P(\Delta \phi)$, for the polar angle difference
between pairs of satellites (diamonds) and \textcolor{black}{the randomized
sample} (circles).  
\textcolor{black}{a)} Full host-satellite sample.  
\textcolor{black}{b)} \textcolor{black}{Satellites of} red hosts.
\textcolor{black}{c)} \textcolor{black}{Satellites of} blue hosts.  
\textcolor{black}{d) Systems with 2 to 3 satellites.
e) Systems with 4 to 6 satellites.
f) Systems with 7 to 26 satellites.}
Values of the reduced $\chi^2$, and the 
probability with which $P(\Delta \phi)$ for the satellites can be 
rejected as having been drawn from
\textcolor{black}{the randomized} distribution, are shown in each panel.
\textcolor{black}{Dashed lines show $P(\Delta \phi)$ for a uniform circular distribution.
Dotted lines in panels a), b), and c) show best-fitting uniform elliptical distributions 
for the observed $P(\Delta \phi)$.}
}
\label{fig:clumping}
\end{figure}

\textcolor{black}{The functional form of $P(\Delta \phi)$ in panels
a) and b) of Fig.\ 4 is similar to that of uniform
elliptical distributions (i.e., the bottom panel of Fig.\ 3).  Motivated
by this, we
compare $P(\Delta \phi)$ for the satellites of all hosts, red hosts,
and blue hosts to $P(\Delta \phi)$ for uniform elliptical 
distributions.  Dotted lines in Fig.\ 4 show best-fit elliptical distributions
in panels a), b), and c).  For the satellites of all hosts and red hosts,
the best-fit elliptical distribution has an axis ratio
$b/a = 0.67$.  For the satellites of blue hosts,
the best-fit elliptical distribution has an axis ratio 
$b/a = 0.70$.  Due to the asymmetric shape of the observed $P(\Delta \phi)$,
the best-fit elliptical distributions
are not good fits ($\chi^2/\nu = 5.7$ for the satellites of
all hosts, $\chi^2/\nu = 1.7$ for the satellites of 
red hosts, and $\chi^2/\nu = 16.0$ for the satellites of
blue hosts).  They are, however, better fits
to the observed $P(\Delta \phi)$ than
are uniform circular distributions.
}


\begin{figure}[t!]
\epsscale{0.75}
\plotone{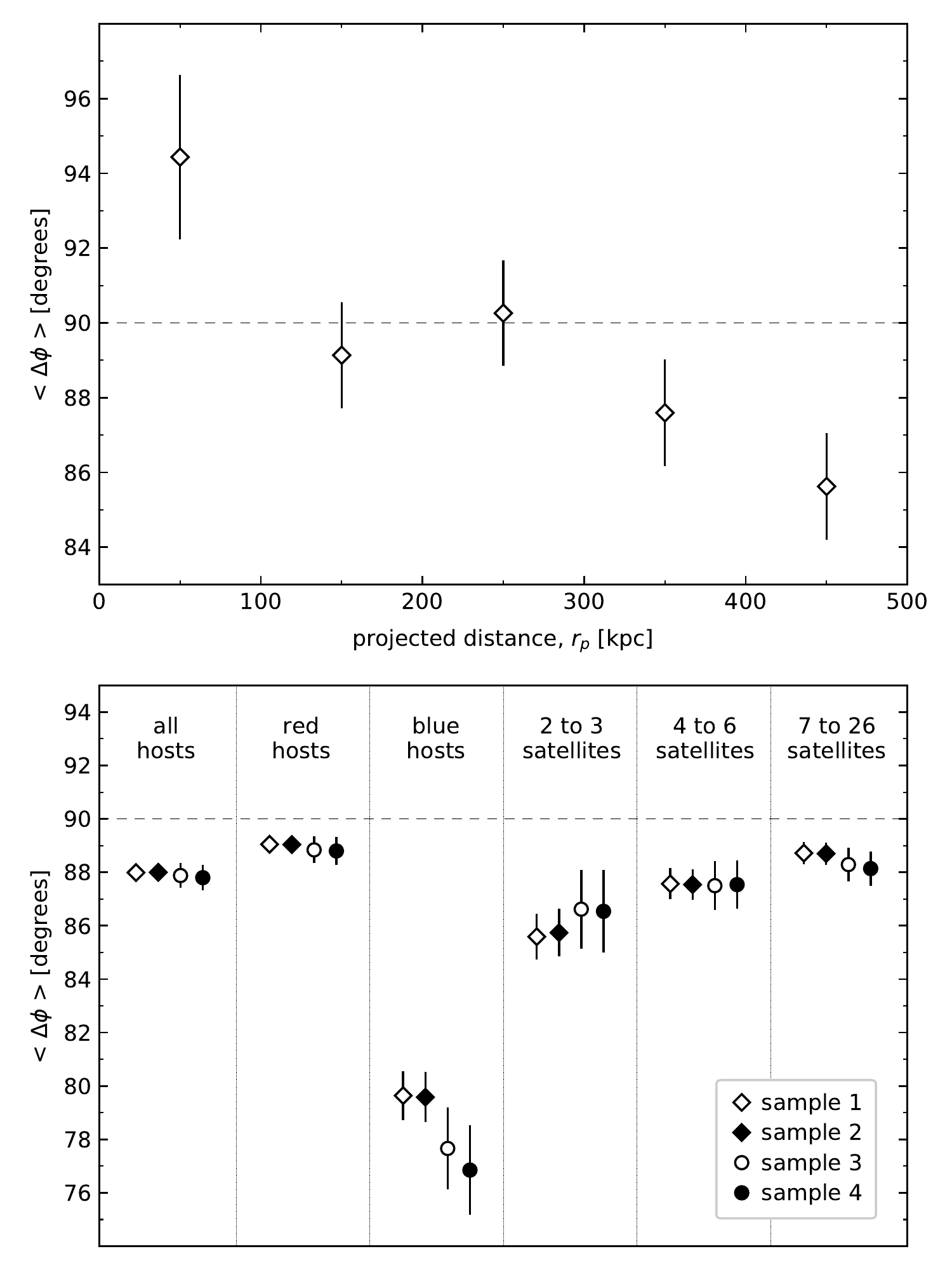}
\caption{
\textcolor{black}{
Mean polar angle difference between pairs of satellites,
$\left< \Delta \phi \right>$.  {\it Top:} Dependence of
$\left< \Delta \phi \right>$ on host-satellite distance in
sample~1.  {\it Bottom:} Dependence of 
$\left< \Delta \phi \right>$ on host isolation criteria for
various subsamples of the host-satellite systems (see text).
}
}
\label{fig:clumping}
\end{figure}

\textcolor{black}{Lastly, we investigate the degree to which 
the lopsidedness of the satellite locations depends on
host-satellite separation and on the isolation criteria 
adopted in sample~1.  To do this,
we quantify the lopsidedness of the satellite
distributions by the
mean polar angle difference between pairs of satellites,
$\left< \Delta \phi \right>$. 
The top panel of Fig.\ 5 shows the dependence
of $\left< \Delta \phi \right>$ on host-satellite separation in 
sample~1.  From this, it is clear that the most distant satellites
($r_p \gtrsim 300$~kpc) are the ones that are responsible for
the overall (i.e., average) tendency for the 
satellites to be found on the same side of their hosts.
Interestingly, there is a weak ($2\sigma$) tendency for
satellites with $r_p \lesssim 100$~kpc to be found
on {\it opposite} sides of their hosts.
}

\textcolor{black}{To investigate the degree to which the
lopsidedness of the satellite locations depends on the
isolation criteria, we 
constructed three additional samples with increasingly strict
isolation criteria.  Sample~2 is the same as sample~1,
except that in sample~2 we reject all host-satellite
systems for which a photometric near neighbor
galaxy exists.  These photometric near neighbors lack spectroscopic redshifts,
but if they happened to be at the same redshift as the host,
they would likely cause the host to fail the 
sample~1 isolation criteria.
These photometric near neighbors
are in the SDSS but not in the NSA, so
their photometry is not identical to that
of the hosts.  For the same galaxy, SDSS magnitudes can be
as much as 0.5~mag fainter than the
NSA magnitudes.  Therefore, we constructed sample~2 by rejecting
any host from sample~1 that was not at least 1.5 magnitudes
brighter than any other galaxy that 
lacked a
spectroscopic redshift and was also located inside the 700~kpc isolation radius.  
Sample~3 was constructed
in the same way as sample~1, except the
isolation radius was increased to 2~Mpc.  Sample~4 was obtained from
sample~3 in the same way that sample~2 was obtained 
from sample~1, except a
2~Mpc isolation radius was used for sample~4.  The bottom panel of Fig.\ 5 shows
a compilation of values of 
$\left< \Delta \phi \right>$ for the various subsamples that we 
analyzed in Fig.\ 4.  From the bottom panel of Fig.\ 5, it is clear that there
is no statistically-significant difference between our results
for the satellite locations in sample~1 and any of
the other samples.  
}

\section{Summary \& Discussion} \label{sec:summary}

We investigated the spatial distribution of the satellites of 
\textcolor{black}{bright, isolated
galaxies,} obtained using typical redshift space criteria.
Our sample differs from previous samples, selected using similar criteria,
in that each of our host-satellite systems \textcolor{black}{is required to
have at least two satellites.}
When averaged over all satellites, the mean satellite
location in our sample, measured with respect to 
the hosts' major axes, agrees well with previous,
\textcolor{black}{similar} studies. 

\textcolor{black}{Since each host-satellite system contains multiple
satellites, we were able to compute
their pairwise clustering.}
The probability distribution for the polar angle
differences between pairs of satellites,
$P(\Delta \phi)$, differs significantly
from \textcolor{black}{$P(\Delta \phi)$ for a randomized distribution}
\textcolor{black}{(confidence level $\ge 99.9999$\%). }
Pairs of satellites show a strong tendency for being
located on the same side of their host, resulting in lopsided 
spatial distributions.  
The effect is most pronounced 
for the satellites of blue hosts \textcolor{black}{
and for systems that
have only 2 or 3 satellites.}
\textcolor{black}{The lopsidedness of the satellite distributions
is robust to substantial changes in the host isolation criteria,
including restricting the sample to 
host galaxies that have no other, similarly-bright galaxy
within a radial distance of 2~Mpc and a line-of-sight
velocity difference $|dv| < 1,000$~km~s$^{-1}$.}

\textcolor{black}{Studies of the satellites of bright galaxy pairs}
(i.e., two similarly bright
galaxies that are relatively close to each other) \textcolor{black}{have
concluded that the spatial distribution of the satellites is lopsided
with respect to the individual host galaxies.}
Ours is the first investigation
to reveal the existence of lopsided \textcolor{black}{distributions for
the satellites of isolated bright galaxies.}
The \textcolor{black}{overall tendency for satellites
to be located on the same side of their host is due to satellites
that are found far from their hosts ($r_p \gtrsim 300$~kpc).  These 
distant satellites are
likely to be recent arrivals to the vicinities of their host
galaxies and are, therefore, unlikely to be a virialized population.
Satellites that are found nearby their hosts ($r_p \lesssim 100$~kpc) 
show a weak preference for being located on opposite
sides of their hosts.  Given that the majority of our host galaxies
have luminosities $L \ge L_\ast$, satellites that are within 100~kpc
of their hosts should be well within the virial radii of their
hosts' dark matter halos.  These nearby satellites
could be expected to be a
virialized population that trace the shapes of their hosts' dark
matter halos, in which case we would expect to find an equal 
number of satellites on the same side of their host as on the 
opposite side.  Considerably larger host-satellite samples will
be needed to determine whether the tendency for 
nearby satellites to be found on opposite sides of their hosts
is statistically-significant in our universe.
}


\acknowledgments

\textcolor{black}{We are deeply grateful to the anonymous reviewer whose
insightful comments and helpful suggestions resulted in substantial
improvements to this {\it Letter}.
}

\end{document}